\documentclass[preprint,aps ,nofootinbib]{revtex4}
\pdfoutput=1
\usepackage{graphicx}
\usepackage{epsfig}
\usepackage{amsmath}
\usepackage{amsfonts}
\usepackage{amssymb}
\usepackage{color}%
\usepackage{dcolumn}
\usepackage{slashed}
\providecommand{\U}[1]{\protect\rule{.1in}{.1in}}

\newcommand{\f}{\begin{equation}}
\newcommand{\ff}{\end{equation}}
\newcommand{\fa}{\begin{eqnarray}}
\newcommand{\ffa}{\end{eqnarray}}

\begin{document}
\title{Holographic fermionic system with dipole coupling on Q-lattice}
\author{Yi Ling $^{1,2}$}
\email{lingy@ihep.ac.cn}
\author{Peng Liu $^{1}$}
\email{liup51@ihep.ac.cn}
\author{Chao Niu $^{1}$}
\email{niuc@ihep.ac.cn}
\author{Jian-Pin Wu $^{3,2}$}
\email{jianpinwu@gmail.com}
\author{Zhuo-Yu Xian $^{1}$}
\email{xianzy@ihep.ac.cn}
\affiliation{$^1$ Institute of High Energy Physics, Chinese Academy of Sciences, Beijing 100049, China\\
$^2$ State Key Laboratory of Theoretical Physics, Institute of Theoretical Physics, Chinese Academy of Sciences, Beijing 100190, China\\
$^3$ Department of Physics, School of Mathematics and Physics, Bohai University, Jinzhou 121013, China}

\begin{abstract}

We construct a holographic model for a fermionic system on
Q-lattice and compute the spectral function in the presence of a
dipole coupling. Both key features of doped Mott insulators, the
dynamical generation of a gap and spectral weight transfer, are
observed when adjusting the value of the coupling parameter $p$.
Of particular interest is that when the background is in a deep
insulating phase, the Mott gap opens much easier with a smaller
coupling parameter in comparison with a metallic background. The
effects of lattice parameters on the width of the gap $\Delta/\mu$
are studied and a turning point is observed near the critical
regime of metal-insulator transitions of the background.
Furthermore, the temperature dependence of the spectral
function is studied.
Finally, we also observe that the anisotropic Q-lattice
generates anisotropic peaks with different magnitudes, indicating that
insulating and metallic phases arise in different directions.

\end{abstract} \maketitle

\section{Introduction}

To understand and describe Mott metal-insulator transition (MIT) is a
long-standing and widely known difficult problem in condensed
matter physics because it involves a strongly correlated electron
system, in which the conventional theoretical tools prove of
little help and non-perturbative techniques are called for. At
this stage, holography may provide insights into the associated
mechanisms of these strongly correlated electron systems by
building a gravitational dual model which is usually solvable in
the large $N$ limit. Some excellent examples are the
holographic superconductor
\cite{Gubser:2008px,Hartnoll:2008vx,Hartnoll:2008kx} and
holographic (non)-Fermi liquid
\cite{SSLee:2008,HongLiu:2009,Zaanen:2009,JPWu:GB,JPWu:Dilaton,WJLiJPWu:Dilaton,XMKuang:2014,Tong:Band,WJLi:Band,LQF:Anisotropic,Reza:1201}.

As early as 1930s, it was reported that many transition-metal
oxides (such as NiO) with partially filled bands show insulating
behavior. Peierls ascribed that to the strong electron-electron
correlation. And then, Mott made a series of pioneering work towards
understanding how electron-electron correlations could explain the
insulating state\cite{Mott1,Mott2,Mott3,Mott4,Mott5}.
Roughly speaking, the main idea of Mott is that
the transition from metal to insulator occurs
as lattice constant increases. Subsequently, this idea of Mott is
formalized in the Hubbard model in which the Mott transition
depends on the competition between the kinetic energy ($E_k\sim
t$) and the on-site Coulomb repulsion $U$. The metallic behavior
prevails if the kinetic energy $t$ overcomes the Coulomb
energy $U$ while insulating phase is favored for $U/t\gg 1$ and a
gap opens in the single-particle excitation spectrum, resulting in
a Mott transition at a critical ratio of $U/t$.

By adding probe fermions with dipole coupling in RN-AdS black
hole\cite{Phillips:PRL,Phillips:PRD}, a Mott gap opens
dynamically, which exhibits two key features of doped Mott
insulator, i.e., the dynamical generation of a gap and spectral
weight transfer. And then, the dipole coupling effects have also
been studied in more general geometries in \cite{Dipole1,Dipole2,Dipole3,Dipole4,Dipole5,Dipole6,WJLi:Dipole,Phillips:SAdS}.
Along this line, here we shall study the holographic fermionic system
with dipole coupling in Q-lattice geometry.

Holographic Q-lattice model is firstly proposed in
\cite{Donos:QLattice}, which is similar to the construction of
Q-balls \cite{Coleman:Qball}. Some extensive studies have been
presented in
\cite{QLattice:Generalize,Donos:Hall,Donos:Ther,YiLingQLatticeS}.
In this framework, the translational symmetry is broken and
MIT is observed through the study of
optical conductivity. Different from the holographic scalar
lattice and ionic lattice constructed in
\cite{Horowitz:Latticev1,Horowitz:Latticev2,Horowitz:Latticev3,YiLing:LatticeFermions,YiLing:Lattice,Donos:1409},
which involve solving the PDEs and need a hard numerical work,
but here we only need solve ODEs for Q-lattice. In particular,
MIT is not observed in scalar lattice
or ionic lattice background yet. The difficulties of numerical
calculation prevent one from dropping the temperature down to an
extremely low level in this context\footnote{MIT is also observed in helical lattice
model\cite{Hartnoll:Nature} and holographic charge density waves\cite{Ling:PRL}.
In addition, a simpler construction which breaks the
translational invariance can be found in
\cite{Withers:2013,Gouteraux:2014}, in which the lattice
amplitude is absent, and thus no MIT happens in these models.}.
Here, we are interested in understanding if there is a relationship between
the phase (metal/insulator) the background geometry with Q-lattice is in
and that the fermionic excitations on Q-lattice are in.
Therefore, in this paper we shall study the fermionic excitations
by adding a probe fermion with dipole coupling in Q-lattice geometry.

This paper is organized as follows. In Section \ref{SQLattices},
based on the holographic model originally presented in
\cite{Donos:QLattice}, which introduces Q-lattice structure in one
spatial direction, we generalize it to a two dimensional lattice
model which in general can be anisotropic in $x-y$ plane. And
then, the Dirac equations over this Q-lattice background are
deduced in Section \ref{SDirac}. In Section \ref{SMott}, we
present our numerical results for the fermionic spectral function.
We conclude in Section \ref{SConclusion} with a summary of main
results and suggestions for future research.

\section{Holographic Q-lattice geometry}\label{SQLattices}

Here we are interested in a holographic Q-lattice geometry with no
translational symmetry in both of the spatial directions. For this
purpose, we consider a system containing two complex scalar fields
$\phi_1$ and $\phi_2$ plus a Maxwell field $A_\mu$. A similar
construction can also be seen in \cite{QLattice:Generalize}.  The
simplest action may take the form as,
\begin{eqnarray}
&&
\label{Action}
S=\int d^4x\sqrt{-g}\left[R+6-\frac{1}{2}F^2-|\partial\phi_1|^2-m_1^2|\phi_1|^2-|\partial\phi_2|^2-m_2^2|\phi_2|^2\right],
\end{eqnarray}
where $F=dA$. The equations of motion can be derived from the
above action as follows
\begin{eqnarray}
&&
\nonumber
R_{\mu\nu}=g_{\mu\nu}(-3+\frac{m_1^2}{2}|\phi_1|^2+\frac{m_2^2}{2}|\phi_2|^2)+\partial_{(\mu}\phi_1\partial_{\nu)}\phi_1^{\ast}+\partial_{(\mu}\phi_2\partial_{\nu)}\phi_2^{\ast}
+(F_{\mu}^{\ \rho}F_{\nu\rho}-\frac{1}{4}g_{\mu\nu}F^2),
\\
\label{EOM}
&&
\nabla_{\mu}F^{\mu\nu}=0,~~~(\nabla^2-m_1^2)\phi_1=0,~~~(\nabla^2-m_2^2)\phi_2=0.
\end{eqnarray}
Consider the following ansatz
\begin{eqnarray}
&&
\nonumber
ds^2=-g_{tt}(z)dt^2+g_{zz}(z)dz^2+g_{xx}(z)dx^2+g_{yy}(z)dy^2,
\\
\
&&
A=A_t(z)dt,~~~~~
\phi_1=e^{ik_1x}\varphi_1,~~~~~
\phi_2=e^{ik_2y}\varphi_2
,
\end{eqnarray}
with
\begin{eqnarray}
&&
\nonumber
g_{tt}(z)=\frac{(1-z)P(z)Q(z)}{z^2},~~~~g_{zz}(z)=\frac{1}{z^2(1-z)P(z)Q(z)},~~
\\
\nonumber
&&
g_{xx}(z)=\frac{V_1(z)}{z^2},~~~~g_{yy}(z)=\frac{V_2(z)}{z^2},~~~~A_t(z)=\mu (1-z)a(z),
\label{Ansatz}
\\
&&
P(z)=1+z+z^2-\frac{\mu^2}{2}z^3
,
\end{eqnarray}
and substitute them into (\ref{EOM}), one has five second order
ODEs for $V_1$, $V_2$, $a$, $\varphi_1$, $\varphi_2$ and one first
order ODE for $Q$. Note that in above ansatz, $k_1$ and $k_2$ are
two wave-numbers along $x$ and $y$ directions, respectively, such
that $\phi_1$ is periodic in $x$ direction with a lattice constant
$2\pi/k_1$ and $\phi_2$ is periodic in $y$ direction with a
lattice constant $2\pi/k_2$. In addition, in our holographic
setup, the dual CFT involves two complex scalar operators with
scaling dimension $\Delta_{1,2}=3/2\pm(9/4+m_{1,2}^2)^{1/2}$. When
either of the masses falls in the interval $[-9/4,-3/2)$, the
$AdS_2$ BF bound near the horizon will be violated by the scalar
field, possibly leading to a different ground state. To avoid this
possibility, we will set $m_{1,2}^2=-5/4$ for definiteness in our
paper, which corresponds to the scaling dimension of
$\Delta_{1,2+}=5/2$ and $\Delta_{1,2-}=1/2$.

To solve the ODEs numerically, we impose a regular boundary
condition at the horizon $z=1$ and impose the following conditions
on the conformal boundary
\begin{eqnarray}
Q(0)=1,~~V_1(0)=V_2(0)=1,~~a(0)=1.
\end{eqnarray}
In this paper, we only focus on the standard quantisation of the
scalar field, in which the asymptotic behaviors of $\varphi_1$ and
$\varphi_2$ look like $\varphi_{1}=\lambda_1 z^{\Delta_{1-}}$ and
$\varphi_{2}=\lambda_2 z^{\Delta_{2-}}$, respectively. The UV
behavior of the scalar field corresponds to a Q-lattice
deformation with lattice amplitudes $\lambda_{1,2}$. In addition,
the temperature of the black hole is given by
\begin{eqnarray}
T=\frac{P(1)Q(1)}{4\pi}.
\end{eqnarray}
As a result, each of our Q-lattice solutions is specified by five
dimensionless quantities $T/\mu$, $\lambda_{1}/\mu^{\Delta_{1-}}$,
$\lambda_{2}/\mu^{\Delta_{2-}}$, $k_{1}/\mu$ and $k_{2}/\mu$. For
simpleness, we shall still use $T$, $\lambda_{1,2}$ and $k_{1,2}$
to denote the above five dimensionless quantities in what follows.

\section {Dirac equation}\label{SDirac}

To explore the properties of fermionic spectral function in the
Q-lattice geometry, we consider the following action including
the dipole coupling with strength $p$ between the fermion and gauge
field\cite{Phillips:PRL,Phillips:PRD,Leighv1,Leighv2}
\begin{eqnarray}
\label{actionspinor}
S_{D}=i\int d^{4}x \sqrt{-g}\overline{\zeta}\left(\Gamma^{a}\mathcal{D}_{a} - m_{\zeta} - ip\slashed{F}\right)\zeta,
\end{eqnarray}
where
$\mathcal{D}_{a}=\partial_{a}+\frac{1}{4}(\omega_{\mu\nu})_{a}\Gamma^{\mu\nu}-iq
A_{a}$ and
$\slashed{F}=\frac{1}{2}\Gamma^{\mu\nu}(e_\mu)^a(e_\nu)^bF_{ab}$
with $(e_{\mu})^{a}$ and $(\omega_{\mu\nu})_{a}$ being a set of
orthogonal normal vector bases and the spin connection 1-forms,
respectively. Here, the fermion is a probe field.
Note that we have set the gauge coupling constant
$g_F=\sqrt{2}$ here. It is different from the conventions in
\cite{Phillips:PRL,Phillips:PRD}, in which they set $g_F=1$.
Consequently the charge $q$ and dipole coupling $p$ here will
correspond to $q/\sqrt{2}$ and $p/\sqrt{2}$ in
\cite{Phillips:PRL,Phillips:PRD} as the relevant quantities are
the products $g_F q$ and $g_F p$.

The Dirac equation can be deduced from the above action
\begin{eqnarray}
\label{DiracEquation1}
\Gamma^{a}\mathcal{D}_{a}\zeta-m_{\zeta}\zeta-ip\slashed{F}\zeta=0.
\end{eqnarray}
To cancel off the spin connection, we can make a redefinition of $\zeta=(g_{tt}g_{xx}g_{yy})^{-\frac{1}{4}}\mathcal{F}$.
At the same time, by the Fourier expansion,
\begin{eqnarray}
\mathcal{F}=\int\frac{d\omega dk_x dk_y}{2\pi}F(z,\textbf{k})e^{-i\omega t + ik_x x + ik_y y},
\end{eqnarray}
where $\textbf{k}=(-\omega,k_x,k_y)$, one has
\begin{eqnarray}
&&
-\frac{1}{\sqrt{g_{zz}}}\Gamma^{3}\partial_{z}F(z,\textbf{k})
+\frac{1}{\sqrt{g_{tt}}}\Gamma^{0}(-i\omega -i q A_{t})F(z,\textbf{k})
+\frac{1}{\sqrt{g_{xx}}}\Gamma^{1} i k_x F(z,\textbf{k})
\nonumber\\
&&
+\frac{1}{\sqrt{g_{yy}}}\Gamma^{2}ik_{y}F(z,\textbf{k})
-m_{\zeta}F(z,\textbf{k})
+\frac{i p}{\sqrt{g_{zz}g_{tt}}}\Gamma^{3}\Gamma^{0} \partial_{z}A_{t} F(z,\textbf{k})
=0.
\end{eqnarray}
Choose the following gamma matrices
\begin{eqnarray}
\label{GammaMatrices}
 && \Gamma^{3} = \left( \begin{array}{cc}
-\sigma^3  & 0  \\
0 & -\sigma^3
\end{array} \right), \;\;
 \Gamma^{0} = \left( \begin{array}{cc}
 i \sigma^1  & 0  \\
0 & i \sigma^1
\end{array} \right),
\cr
&&
\Gamma^{1} = \left( \begin{array}{cc}
-\sigma^2  & 0  \\
0 & \sigma^2
\end{array} \right), \;\;
 \Gamma^{2} = \left( \begin{array}{cc}
 0  & \sigma^2  \\
\sigma^2 & 0
\end{array} \right),
\end{eqnarray}
and split the 4-component spinor into two 2-component spinors as $F=(F_{1},F_{2})^{T}$, one has
\begin{eqnarray}
\label{DiracETotalv1}
&&
\frac{1}{\sqrt{g_{zz}}}\partial_{z} \left( \begin{matrix} F_{1}(\textbf{k}) \cr  F_{2}(\textbf{k}) \end{matrix}\right)
-m_{\zeta}\sigma^3\otimes\left( \begin{matrix} F_{1}(\textbf{k}) \cr  F_{2}(\textbf{k}) \end{matrix}\right)
+(\omega+ q A_{t})\frac{1}{\sqrt{g_{tt}}}i\sigma^2\otimes \left( \begin{matrix} F_{1}(\textbf{k}) \cr  F_{2}(\textbf{k}) \end{matrix}\right)
\nonumber
\\
&&
\mp k_x \frac{1}{\sqrt{g_{xx}}} \sigma^1 \otimes \left( \begin{matrix} F_{1}(\textbf{k}) \cr  F_{2}(\textbf{k}) \end{matrix}\right)
+ \frac{k_y}{\sqrt{g_{yy}}}\sigma^1\otimes \left( \begin{matrix} F_{2}(\textbf{k}) \cr  F_{1}(\textbf{k}) \end{matrix}\right)
+\frac{p}{\sqrt{g_{zz}g_{tt}}}(\partial_{z}A_{t})\sigma^1 \otimes \left( \begin{matrix} F_{1}(\textbf{k}) \cr  F_{2}(\textbf{k}) \end{matrix}\right)
\nonumber
\\
&&
=0.
\end{eqnarray}
Furthermore, by the decomposition
$
F_{\alpha} \equiv (\mathcal{A}_{\alpha}, \mathcal{B}_{\alpha})^{T}
$
with $\alpha=1,2$, the above Dirac equation can be expressed as
\begin{eqnarray} \label{DiracEAB1}
&&
\left(\frac{1}{\sqrt{g_{zz}}}\partial_{z}\mp m_{\zeta} \right)\left( \begin{matrix} \mathcal{A}_{1} \cr  \mathcal{B}_{1} \end{matrix}\right)
\pm (\omega+ q A_{t})\frac{1}{\sqrt{g_{tt}}}\left( \begin{matrix} \mathcal{B}_{1} \cr  \mathcal{A}_{1} \end{matrix}\right)
+\frac{p}{\sqrt{g_{zz}g_{tt}}}(\partial_{z}A_{t})\left( \begin{matrix} \mathcal{B}_{1} \cr  \mathcal{A}_{1} \end{matrix}\right)
\nonumber
\\
&&
- \frac{k_x}{\sqrt{g_{xx}}} \left( \begin{matrix} \mathcal{B}_{1} \cr  \mathcal{A}_{1} \end{matrix}\right)
+ \frac{k_y}{\sqrt{g_{yy}}} \left( \begin{matrix} \mathcal{B}_{2} \cr  \mathcal{A}_{2} \end{matrix}\right)
=0,
\end{eqnarray}
\begin{eqnarray} \label{DiracEAB2}
&&
\left(\frac{1}{\sqrt{g_{zz}}}\partial_{z}\mp m_{\zeta}\right)\left( \begin{matrix} \mathcal{A}_{2}\cr  \mathcal{B}_{1}\end{matrix}\right)
\pm (\omega+ q A_{t})\frac{1}{\sqrt{g_{tt}}}\left( \begin{matrix} \mathcal{B}_{2}\cr  \mathcal{A}_{2}\end{matrix}\right)
+\frac{p}{\sqrt{g_{zz}g_{tt}}}(\partial_{z}A_{t})\left( \begin{matrix} \mathcal{B}_{2} \cr  \mathcal{A}_{2} \end{matrix}\right)
\nonumber
\\
&&
+\frac{k_x}{\sqrt{g_{xx}}} \left( \begin{matrix} \mathcal{B}_{2}\cr  \mathcal{A}_{2}\end{matrix}\right)
+ \frac{k_y}{\sqrt{g_{yy}}} \left( \begin{matrix} \mathcal{B}_{1}\cr  \mathcal{A}_{1}\end{matrix}\right)
=0.
\end{eqnarray}
At the horizon, we can find that
\begin{eqnarray}
\label{DiracEABalphaNearHorizon1}
&&
\partial_{z}\left( \begin{matrix} \mathcal{A}_{\alpha}(z,\textbf{k}) \cr  \mathcal{B}_{\alpha}(z,\textbf{k}) \end{matrix}\right)
\pm \frac{\omega}{4\pi T}\frac{1}{1-z}
\left( \begin{matrix} \mathcal{B}_{\alpha}(z,\textbf{k}) \cr  \mathcal{A}_{\alpha}(z,\textbf{k}) \end{matrix}\right)
=0.
\end{eqnarray}
In order to obtain the retarded Green function on
the boundary by holography, the independent ingoing boundary
condition should be imposed at the horizon, i.e.,
\begin{equation}
\left( \begin{matrix} \mathcal{A}_{\alpha}(z,\textbf{k}) \cr  \mathcal{B}_{\alpha}(z,\textbf{k}) \end{matrix}\right)
=c_\alpha\left( \begin{matrix} 1
\cr  -i\end{matrix}\right)(1-z)^{-\frac{i\omega}{4\pi T}}.
\end{equation}
Near the AdS boundary, the Dirac field reduces to
\begin{eqnarray} \label{BoundaryBehaviour}
\left( \begin{matrix} \mathcal{A}_{\alpha} \cr  \mathcal{B}_{\alpha}\end{matrix}\right)
{\approx} a_{\alpha}z^{m_{\zeta}}\left( \begin{matrix} 1 \cr  0 \end{matrix}\right)
+b_{\alpha}z^{-m_{\zeta}}\left( \begin{matrix} 0 \cr 1 \end{matrix}\right).
\end{eqnarray}
And so by holography, the retarded Green function can be read off
\begin{eqnarray}
a_{\alpha}=G_{\alpha\alpha'}b_{\alpha'}.
\end{eqnarray}
Note that since the four components of the Dirac fields couple to
one another, we need to construct a basis of finite solutions,
$(\mathcal{A}^{I}_{\alpha},\mathcal{B}^{I}_{\alpha})$ and
$(\mathcal{A}^{II}_{\alpha},\mathcal{B}^{II}_{\alpha})$, to obtain
the boundary Green function. We are mainly interested in the
measurable spectral function, which is $A(\omega,k_x,k_y)\sim Im(Tr
G)$.

\section{Mott transition on Q-lattice}\label{SMott}

In this section, we study the properties of spectral function on
Q-lattice. We shall firstly address Mott transition in an
isotropic Q-lattice, in which we set $\lambda_1=\lambda_2$ and
$k_1=k_2$. So, in this paper, we shall denote
$\lambda=\lambda_1=\lambda_2$ and $k=k_1=k_2$ except in Subsection
\ref{anisotropic}, where we give a brief discussion on anisotropic
spectral function. For definiteness, we work exclusively with
the massless fermion and fix $q=1$. In addition, we
work at a very low but non-zero temperature of $T\simeq 0.00398$ for this paper except the subsection \ref{ST}, in which we
explore the dynamics at different temperatures.

\subsection{Free fermionic spectral function}\label{SuSSFWiDipole}

\begin{figure}
\center{
\includegraphics[scale=0.25]{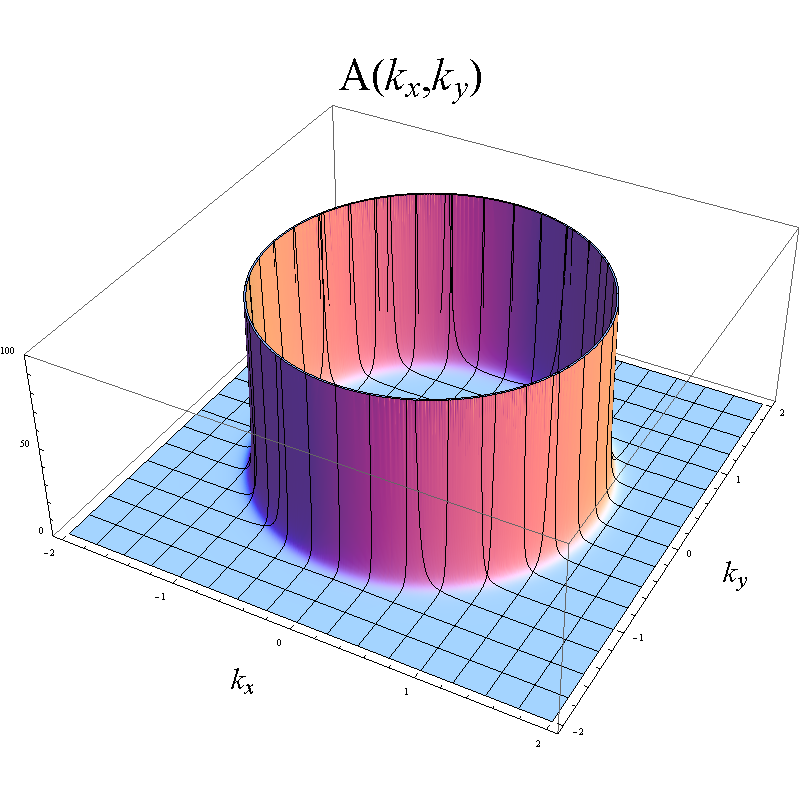}\hspace{0.5cm}
\includegraphics[scale=0.25]{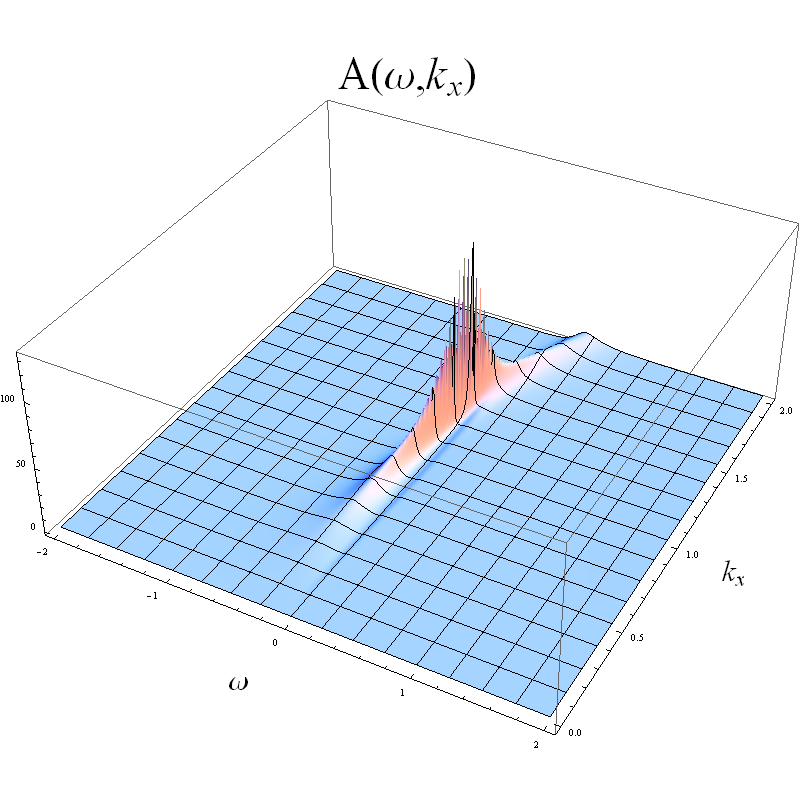}\hspace{0.2cm}
\caption{\label{D3v1}Left plot: The spectral function $A(k_x,k_y)$ for a tiny $\omega$.
Right plot: The spectral function $A(\omega,k_x)$ for $k_y=0$.
Here, we have fixed $\lambda=0.5$ and $k=0.8$.}}
\end{figure}

In this subsection, we present the results for free fermionic
spectral function on Q-lattice. Because the notion of
Fermi surface is only well defined at zero temperature, for our
Q-lattice system of low but non-zero temperature, we need an
operational definition, which has been proposed in
\cite{FermiSurface:Nature} and adopted in holographic models
\cite{Herzog:FermiArcs,YiLing:LatticeFermions}. It is argued that
the Fermi momentum $k_F$ can be determined by searching the peak
of $A(\omega,k)$ with a tiny frequency $\omega$. With the use of
this operational definition, we show the 3D plot of spectral
function $A(k_x,k_y)$ for a tiny $\omega$ in the left plot in
FIG.\ref{D3v1}. We can see that the shape of Fermi surface is a
circle, which is in agreement with the fact that our holographic
Q-lattice geometry is isotropic. For convenience, we shall set
$k_y=0$ in what follows. As an example, the 3D plot of
$A(\omega,k_x)$ is shown in the right plot in FIG.\ref{D3v1}. A
peak occurs near $\omega=0$ and $k_F\simeq 1.359$, indicating a
Fermi surface.

\begin{figure}
\center{
\includegraphics[scale=0.6]{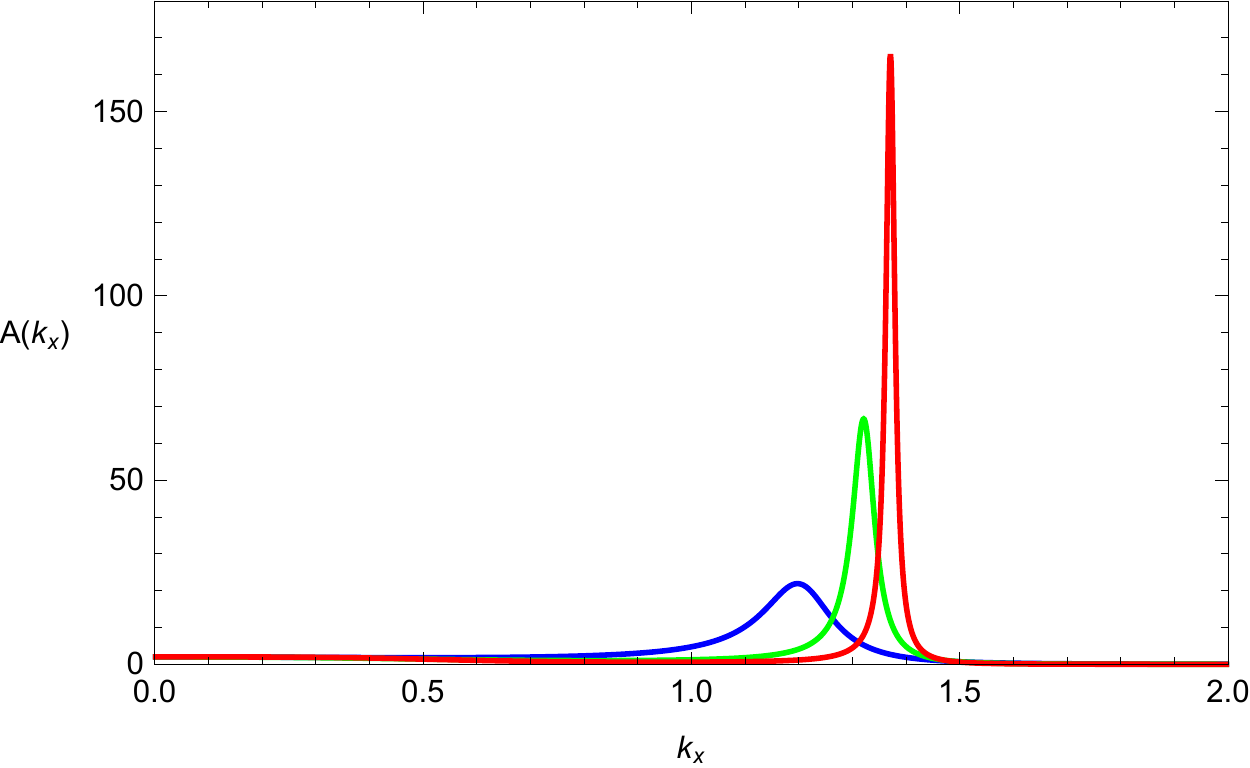}\hspace{0.2cm}
\includegraphics[scale=0.6]{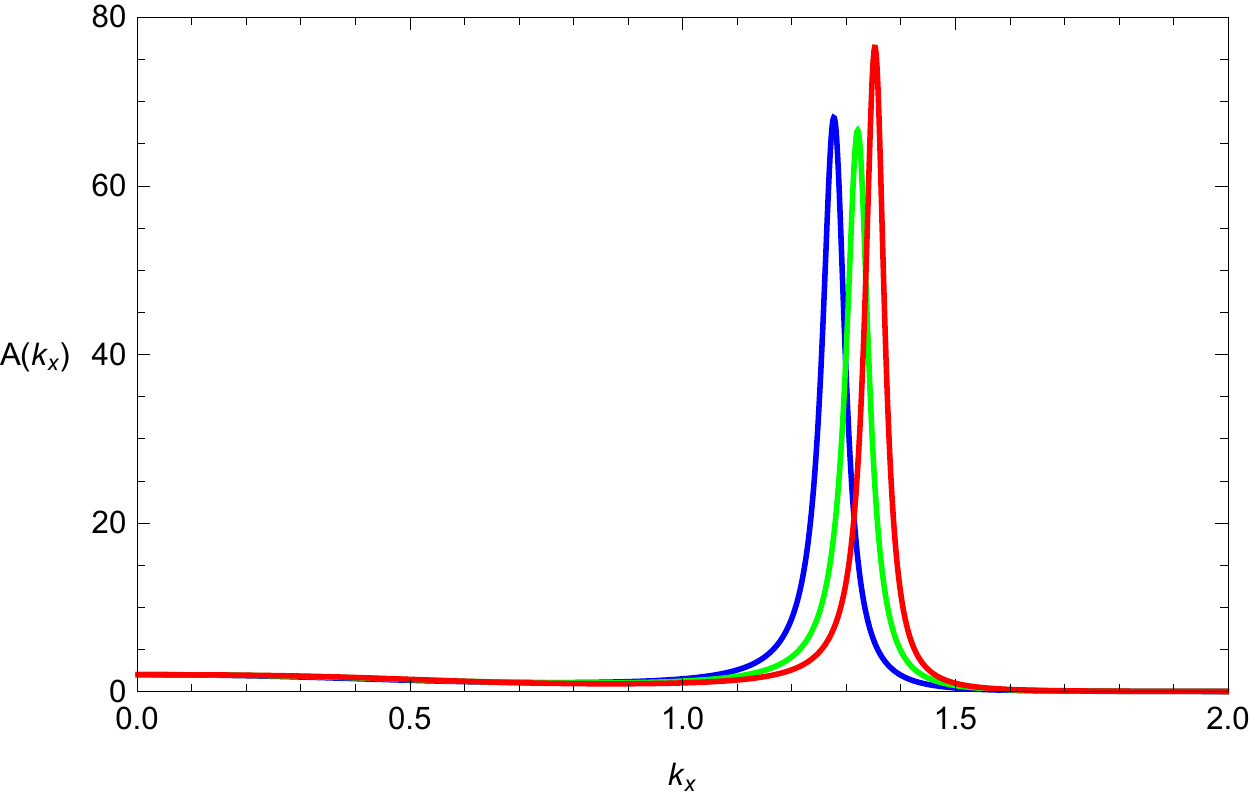}\hspace{0.2cm}
\caption{\label{AvkDlambda}Left plot: The spectral function $A(k_x)$ for fixed $k=0.8$ and different
$\lambda$ (red line for $\lambda=0.1$, green for $\lambda=1$ and blue for $\lambda=2$).
Right plot: The spectral function $A(k_x)$ for fixed $\lambda=1$ and different
$k$ (red line for $k=1$, green for $k=0.8$ and blue for $k=0.1$). Here, all plots are for $k_y=0$ and a tiny $\omega$.}}
\end{figure}

The Q-lattice effects on the evolution of the free fermionic
spectral function can be seen as follows. On one hand, one can fix
the wave-number and see the changes of spectrum with the lattice
amplitude. We show an example on the left plot in
FIG.\ref{AvkDlambda}, where $k=0.8$. For large $\lambda$
($\lambda=2$, blue line in the left plot in FIG.\ref{AvkDlambda}),
only a small bump is displayed. With the decrease of $\lambda$,
the small bump grows into a peak, which indicates the emergence of
the Fermi surface. On the other hand, one can fix the lattice
amplitude and adjust the wave-number $k$ to see its impact on the
shape and location of Fermi surface. From the right plot in
FIG.\ref{AvkDlambda}, we see that with the augmentation of
the wave-number $k$, the Fermi peak shifts to positions with
larger momenta. Moreover, we notice that the height of peak for
$k=0.8$ is lower than that for $k=0.1$ or $k=1$. It appears to be
a turning point around $k=0.8$. In the subsequent subsection, such
a turning point can be further confirmed by studying
the critical value $p_c$ of the onset of the gap and the
gap width $\Delta/\mu$.

Finally, we would like to point out that
no matter how to tune the
lattice amplitude $\lambda$ and wave-number $k$, one can not
observe a gap open at the Fermi level ($\omega\simeq 0$), even
though the Q-lattice background is dual to a deep insulating
phase.
It implies that in the absence of coupling, the holographic
Q-lattice background itself is not able to drive fermionic probes
to undergo a Mott transition. To model Mott physics, we shall
introduce the dipole coupling term as proposed in
\cite{Phillips:PRL}.

\subsection{The Mott transition}

\begin{figure}
\center{
\includegraphics[scale=0.25]{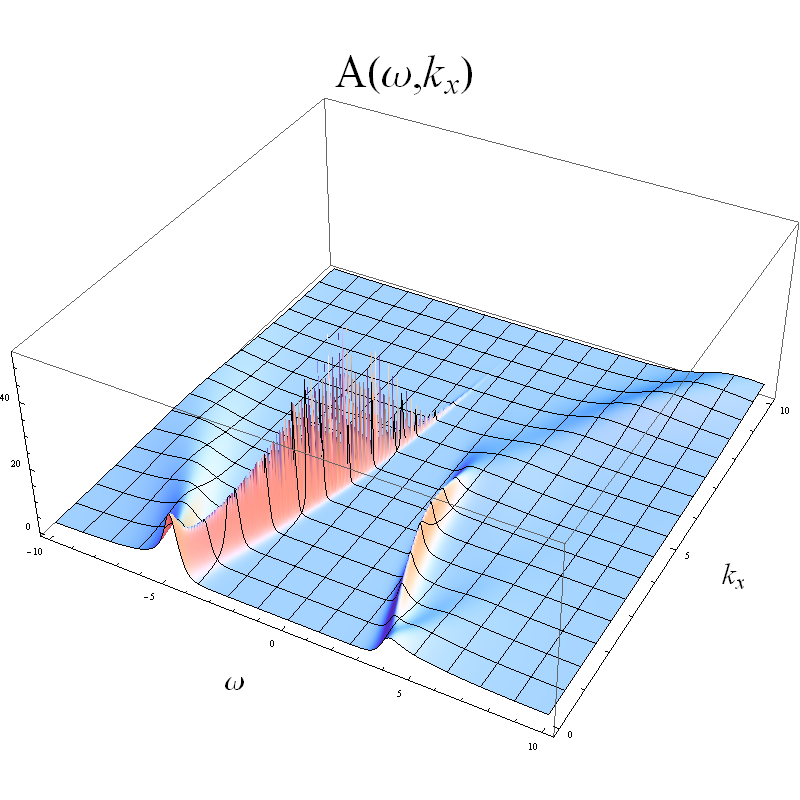}\hspace{0.2cm}
\caption{\label{Gap_D3}The spectral function $A(\omega,k_x)$ for $p=4.5$. Here, $\lambda=0.5$, $k=0.8$ and $k_y=0$.
A gap is clearly visible around $\omega=0$.}}
\end{figure}
\begin{figure}
\center{
\includegraphics[scale=0.6]{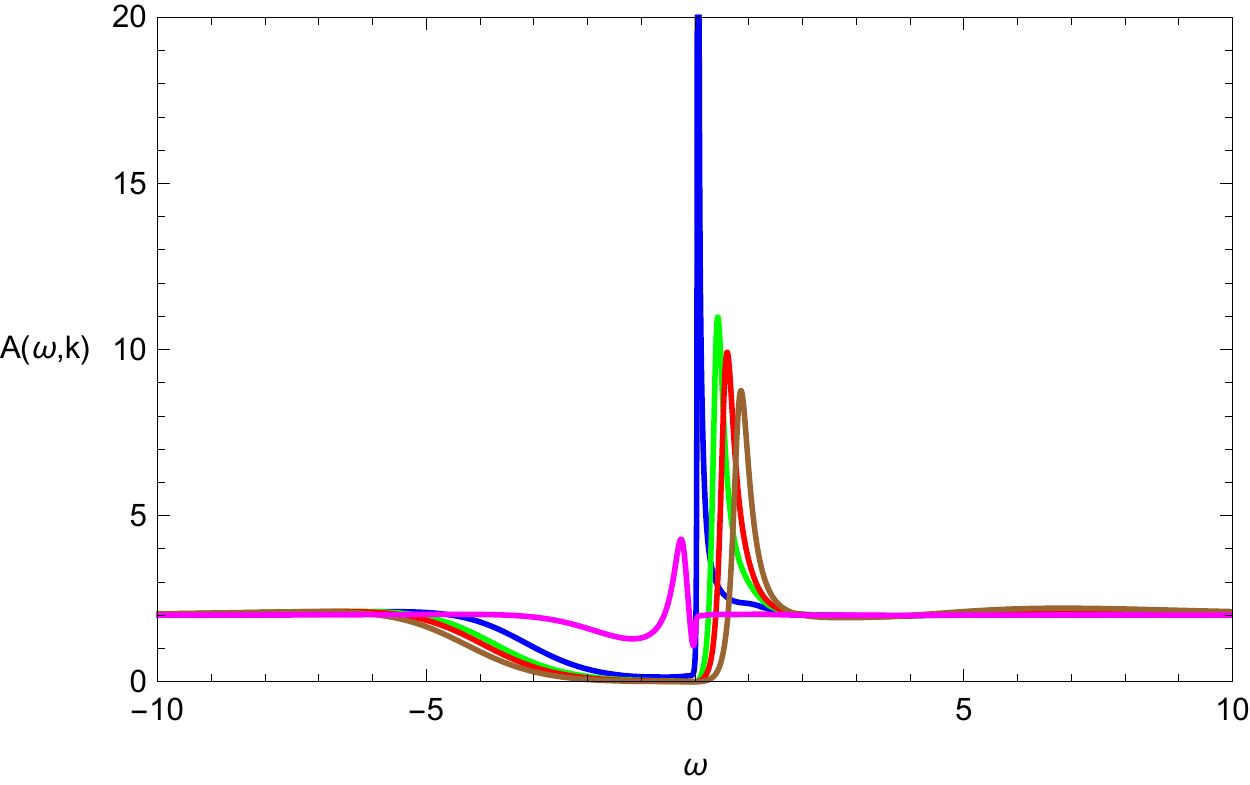}\hspace{0.2cm}
\includegraphics[scale=0.6]{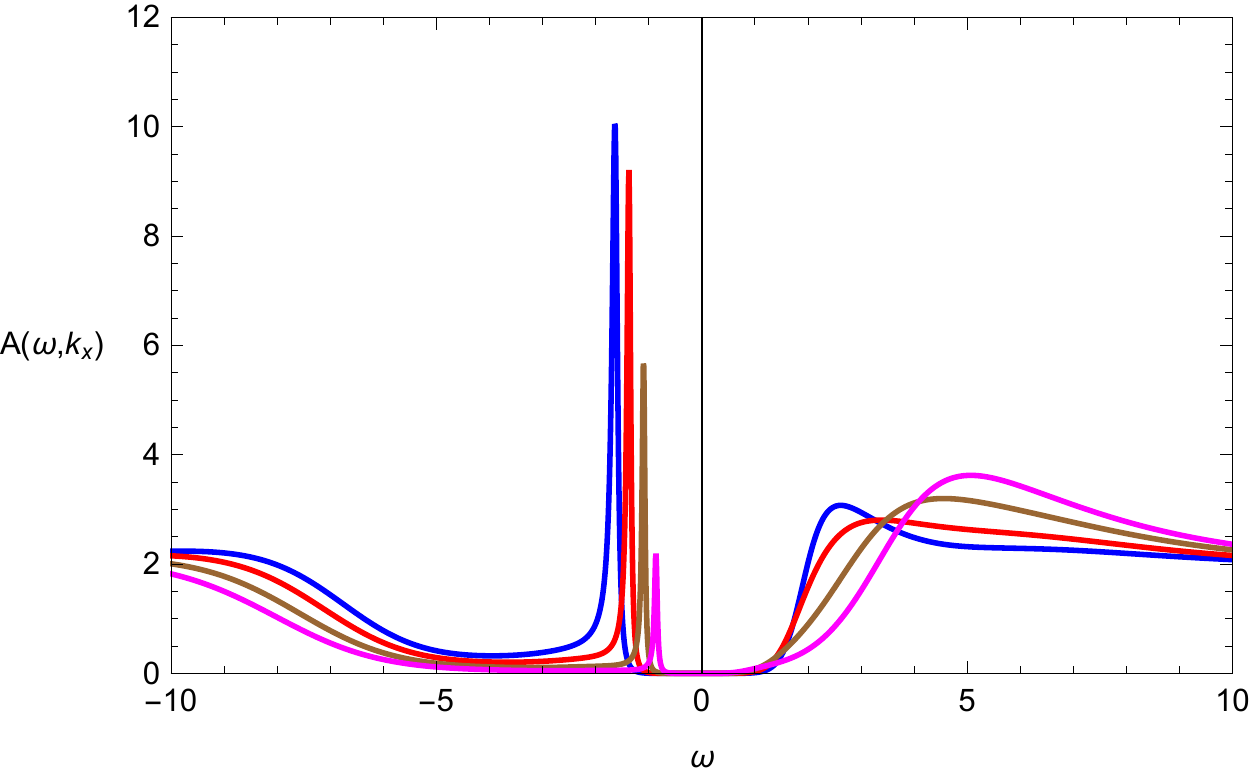}\hspace{0.2cm}
\caption{\label{AvomegaFork}The spectral function $A(\omega,k_x)$ for sample values of $k_x$.
Left plot is for $p=0$ and right for $p=4.5$. Here, $\lambda=0.5$, $k=0.8$ and $k_y=0$.}}
\end{figure}

A Mott insulator is signaled by a vanishing spectral weight, i.e.,
a suppression of $A(\omega)$ around $\omega=0$. The 3D plot of
spectral function $A(\omega,k_x)$ with $p=4.5$ for $\lambda=0.5$
and $k=0.8$ in FIG.\ref{Gap_D3} shows such a gap, which has
previously been observed in RN-AdS geometry
\cite{Phillips:PRL,Phillips:PRD} and other backgrounds
\cite{Dipole1,Dipole2,Dipole3,Dipole4,Dipole5,Dipole6}. In
addition, two bands are located in the positive frequency and
negative frequency regions, respectively. Furthermore, we also
show the spectral function $A(\omega,k_x)$ with fixed $p$ for
sample values of $k_x$ in FIG.\ref{AvomegaFork}. For $p=0$ (the left
plot in FIG.\ref{AvomegaFork}), as the peak approaches $\omega=0$,
it becomes sharper and narrower, indicating a Fermi surface at
$k=k_F$. While the right plot in FIG.\ref{AvomegaFork} shows the
spectral function $A(\omega,k_x)$ with $p=4.5$ for sample values
of $k_x$. It indicates that when $p$ 
exceeds
certain critical point $p_c$, a gap opens and exists for all
$k_x$. The above observation strongly manifests that a transition
occurs from a (non)-Fermi like phase to an insulating phase.

\begin{figure}
\center{
\includegraphics[scale=0.6]{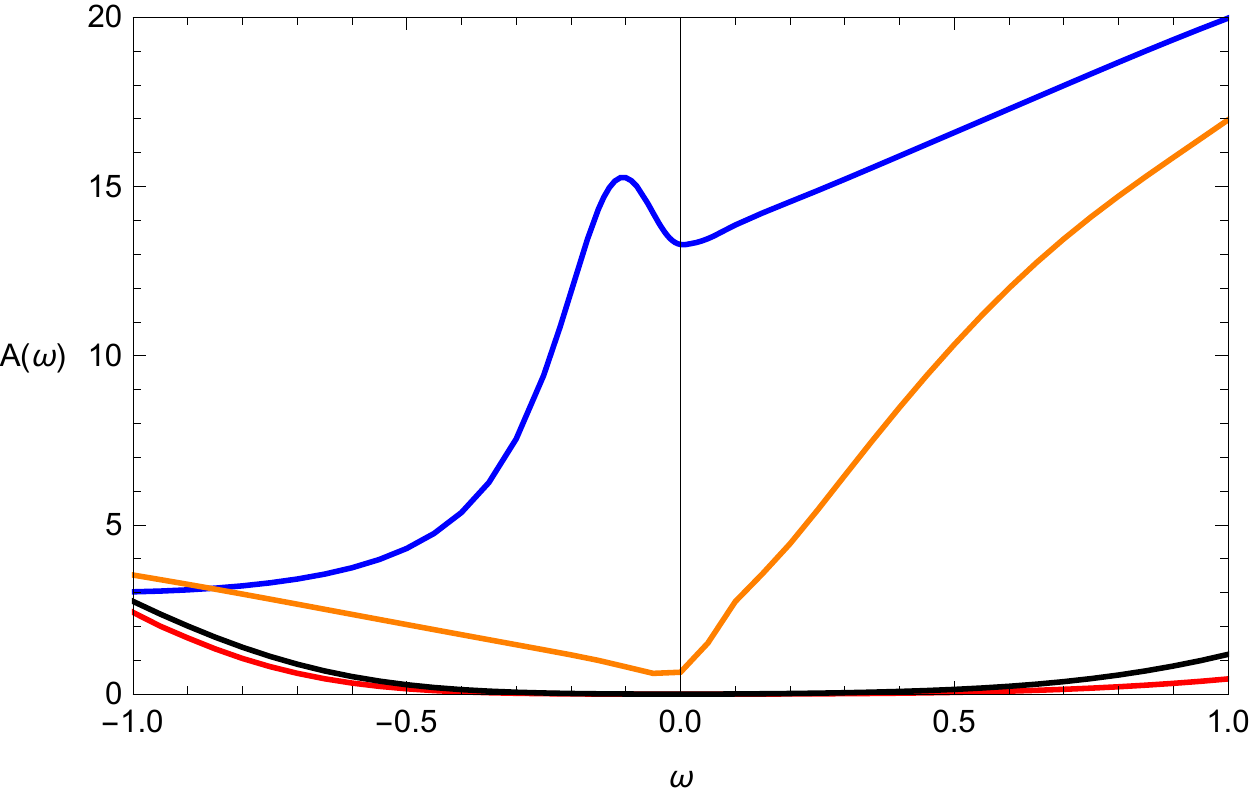}\hspace{0.2cm}
\caption{\label{DOSVp}The DOS $A(\omega)$ with $\lambda=0.5$ and $k=0.8$ for $p=0$ (blue), $p=1.5$ (orange), $p=4$ (black) and $p=4.5$ (red).
}}
\end{figure}

To gain a better understanding of the spectral measure with dipole
coupling on Q-lattice, we study the density of states (DOS) for our
fermionic system. The DOS $A(\omega)$ signals the total weight of
the spectral measure and is defined by the integral of
$A(\omega,k_x,k_y)$ over $k_x$ and $k_y$. Since we have set
$k_y=0$, the DOS only probes the effects in $k_x$ direction.
FIG.\ref{DOSVp} shows the DOS $A(\omega)$ with $\lambda=0.5$ and
$k=0.8$, and exhibits the dynamical generation of a gap with the
increase of the diploe coupling $p$. More importantly, as $p$
increases, the spectral weight switches gradually from the
positive frequency region to the negative frequency region. The
dynamical generation and spectral weight transfer are two key
features of doped Mott
physics. Such observations are confirmed by checking 
other values of lattice parameters $\lambda$ and $k$. Therefore,
the Mott transition in holographic Q-lattice fermionic system can
be implemented by introducing a dipole coupling, which is supposed
to play double roles of doping as well as on-site
interaction like $U$ in Hubbard model, as argued in RN-AdS
geometry \cite{Phillips:PRL,Phillips:PRD}.

Now, we turn to study the critical value $p_c$ of Mott transition,
which reflects how easy it is to open the gap dynamically. In
numerical calculations, the onset of gap can be identified as that
the DOS $A(\omega)$ at the Fermi level drops below some small
number (here, we take $10^{-3}$). With this in mind, we find that
the critical value $p_c\simeq 3.98$ for $\lambda=0.5$ and $k=0.8$,
which corresponds to a metallic phase in \cite{Donos:QLattice}.
While for an insulating phase with $\lambda=2$ and $k=1/2^{3/2}$,
$p_c=3.74$. It implies that the gap opens easier in a deep
insulating phase than metallic phase. This observation can be
further confirmed by evaluating the critical value for various
wave number $k$ and lattice amplitude $\lambda$, as we list in
Table \ref{Tablev1} and \ref{Tablev2}. Nevertheless, we notice
that an anomalous behavior occurs for $k\geq k_c$ ( where $k_c$ is
the critical value for MIT at zero
temperature, with $k_c\simeq 0.8$ for $\lambda=1$), in which the
critical value $p_c$ diminishes slightly with the increase of $k$,
as opposite to that for $k<k_c$.

\begin{figure}
\center{
\includegraphics[scale=0.6]{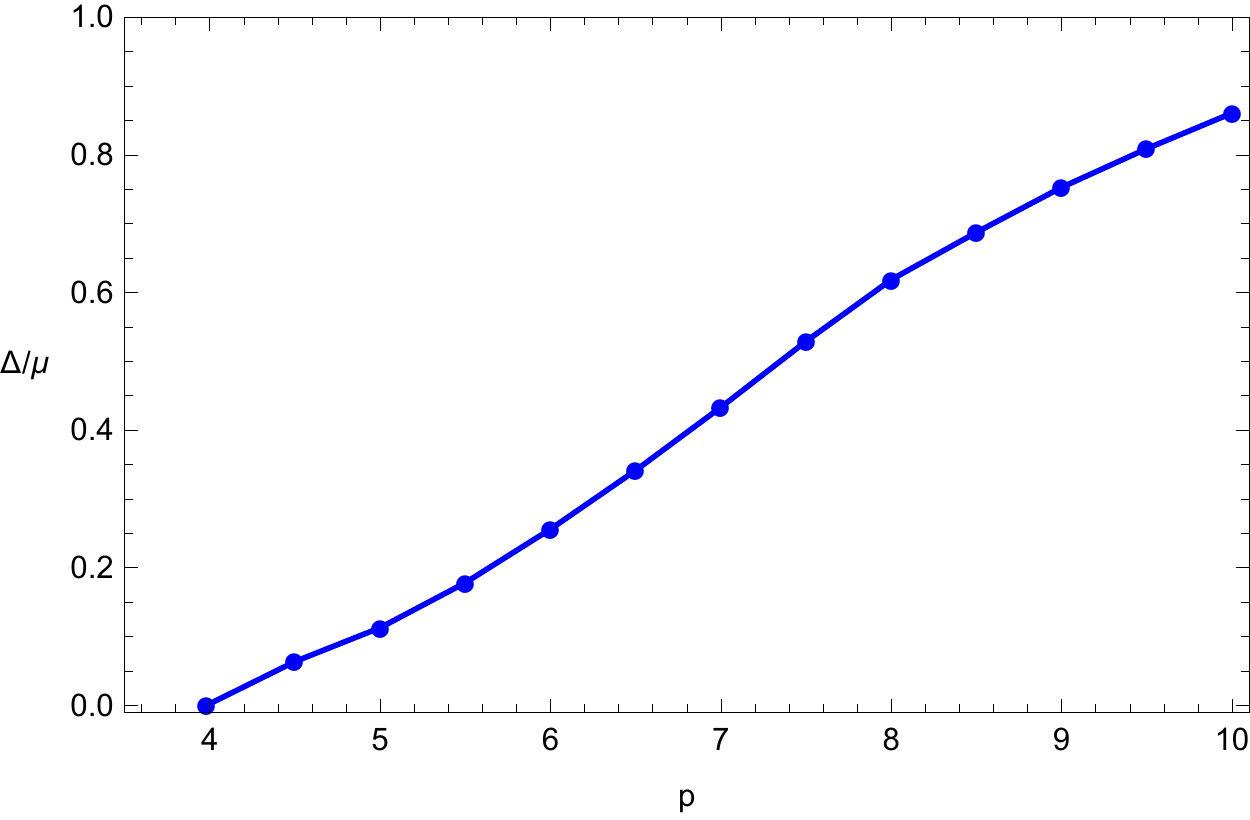}\hspace{0.5cm}
\caption{\label{widthp}The width $\Delta/\mu$ of the gap as a function of the dipole coupling $p$ for $\lambda_1=\lambda_2=0.5$ and $k_1=k_2=0.8$.}}
\end{figure}

\begin{widetext}
\begin{table}[ht]
\begin{center}
\begin{tabular}{|c|c|c|c|c|c|c|}
         \hline
~$k$~ &~$0.1$~&~$0.3$~&~$0.5$~&~$0.8$~&~$1$~&~$1.5$~
          \\
        \hline
~$p_c$~ & ~$3.44$~ & ~$3.76$~ & ~$4.08$~&~$4.12$~&~$4.08$~&~$3.93$~
          \\
        \hline
\end{tabular}
\caption{\label{Tablev1}$p_c$ for sample values of $k$ and fixed $\lambda=1$.}
\end{center}
\end{table}
\end{widetext}
\begin{widetext}
\begin{table}[ht]
\begin{center}
\begin{tabular}{|c|c|c|c|c|c|c|}
         \hline
~$\lambda$~ &~$0.1$~&~$0.5$~&~$1$~&~$1.5$~&~$2$~&~$3$~
          \\
        \hline
~$p_c$~ & ~$3.90$~ & ~$3.83$~ & ~$3.76$~&~$3.7$~&~$3.64$~&~$3.55$~
          \\
        \hline
\end{tabular}
\caption{\label{Tablev2}$p_c$ for sample values of $\lambda$ and fixed $k=0.3$.}
\end{center}
\end{table}
\end{widetext}

Subsequently, we shall study the features of Mott insulating phase
over a Q-lattice background. A characteristic measure is the width
of the gap $\Delta/\mu$. FIG.\ref{widthp} shows the width
$\Delta/\mu$ as a function of the dipole coupling $p$ for
$\lambda=0.5$ and $k=0.8$. As $p$ increases, the width of the gap
becomes larger. It is just the effects of on-site interaction
strength $U$ of Hubbard model, which $p$ plays.

\begin{figure}
\center{
\includegraphics[scale=0.6]{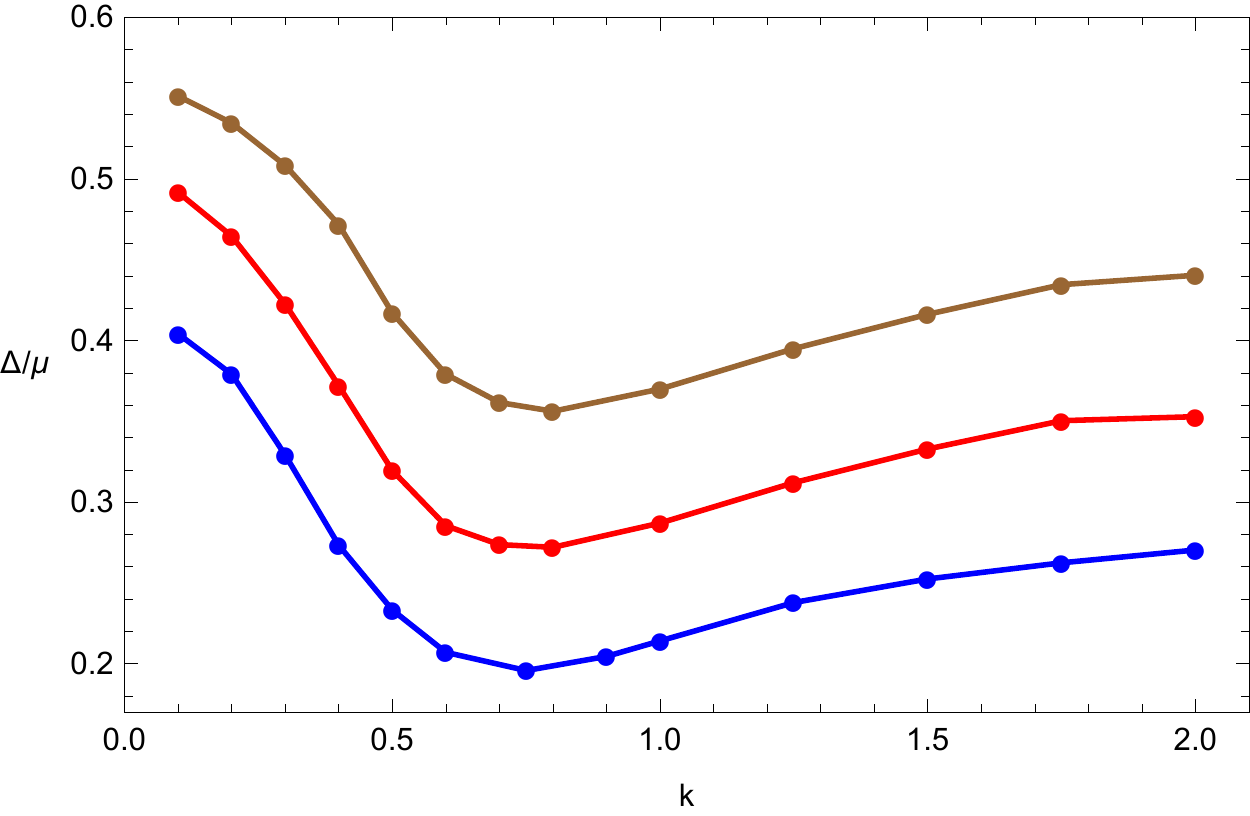}\hspace{0.5cm}
\includegraphics[scale=0.6]{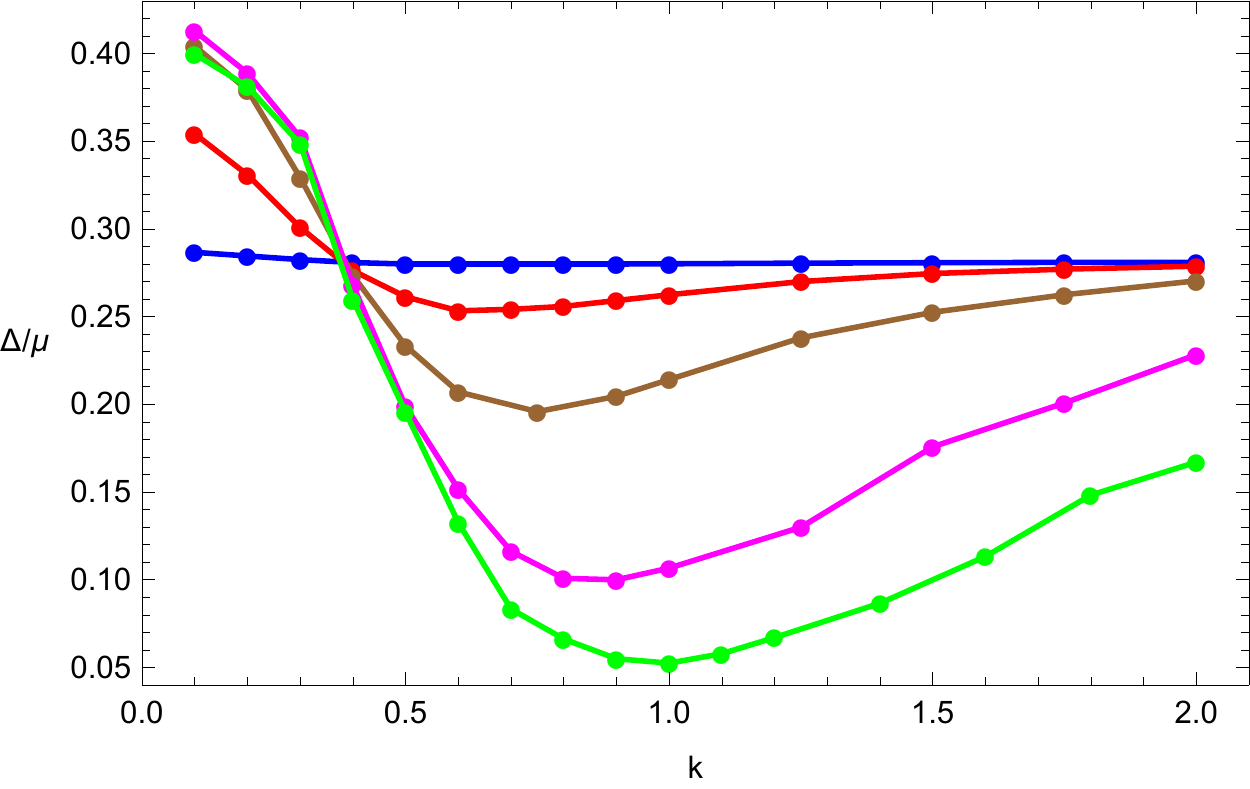}\hspace{0.5cm}
\caption{\label{width_k_lambda}The width $\Delta/\mu$ of the gap as a function of $k$.
Left plot is for $\lambda=1$ and different $p$ (blue line for $p=6$, red for $p=6.5$ and brown for $p=7$)
and right plot for $p=6$ and different $\lambda$ (blue line for $\lambda=0.1$, red for $\lambda=0.5$, brown for $\lambda=1$, magnetic for $\lambda=2$ and green for $\lambda=3$).}}
\end{figure}

\begin{figure}
\center{
\includegraphics[scale=0.6]{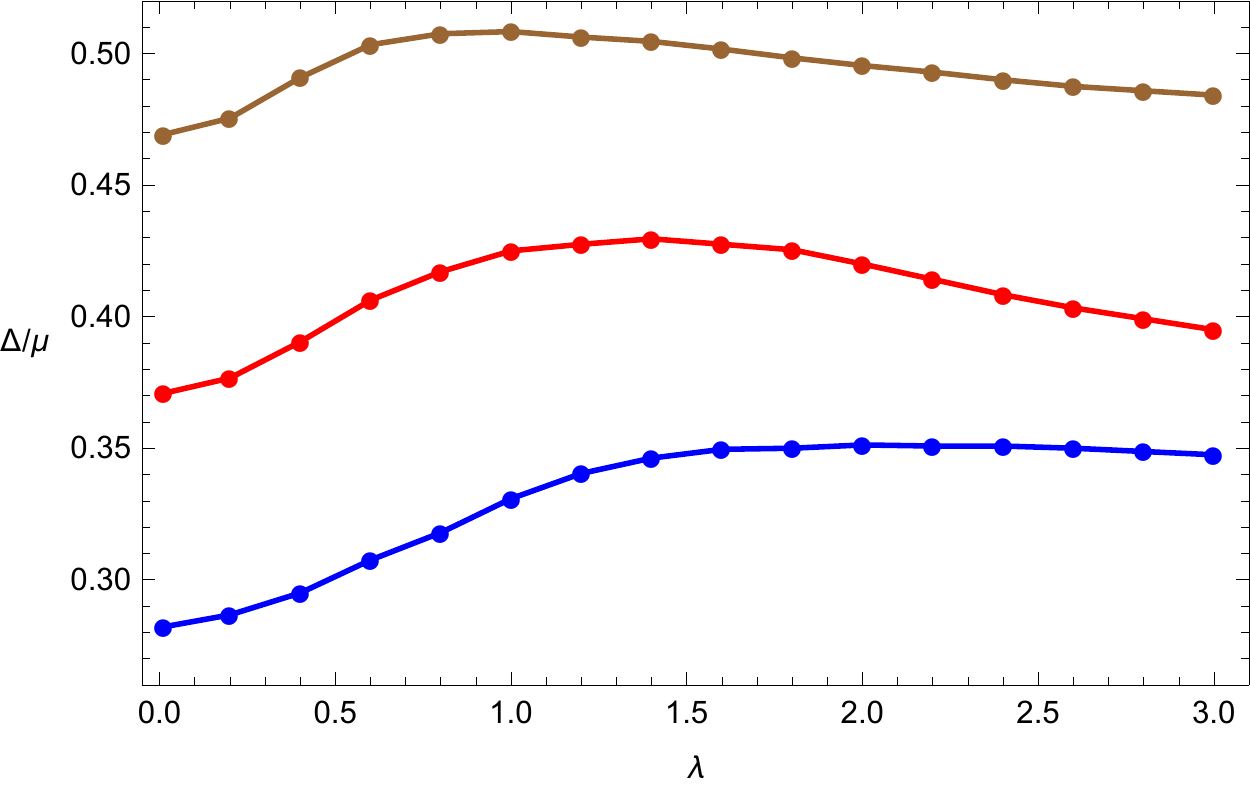}\hspace{0.5cm}
\includegraphics[scale=0.6]{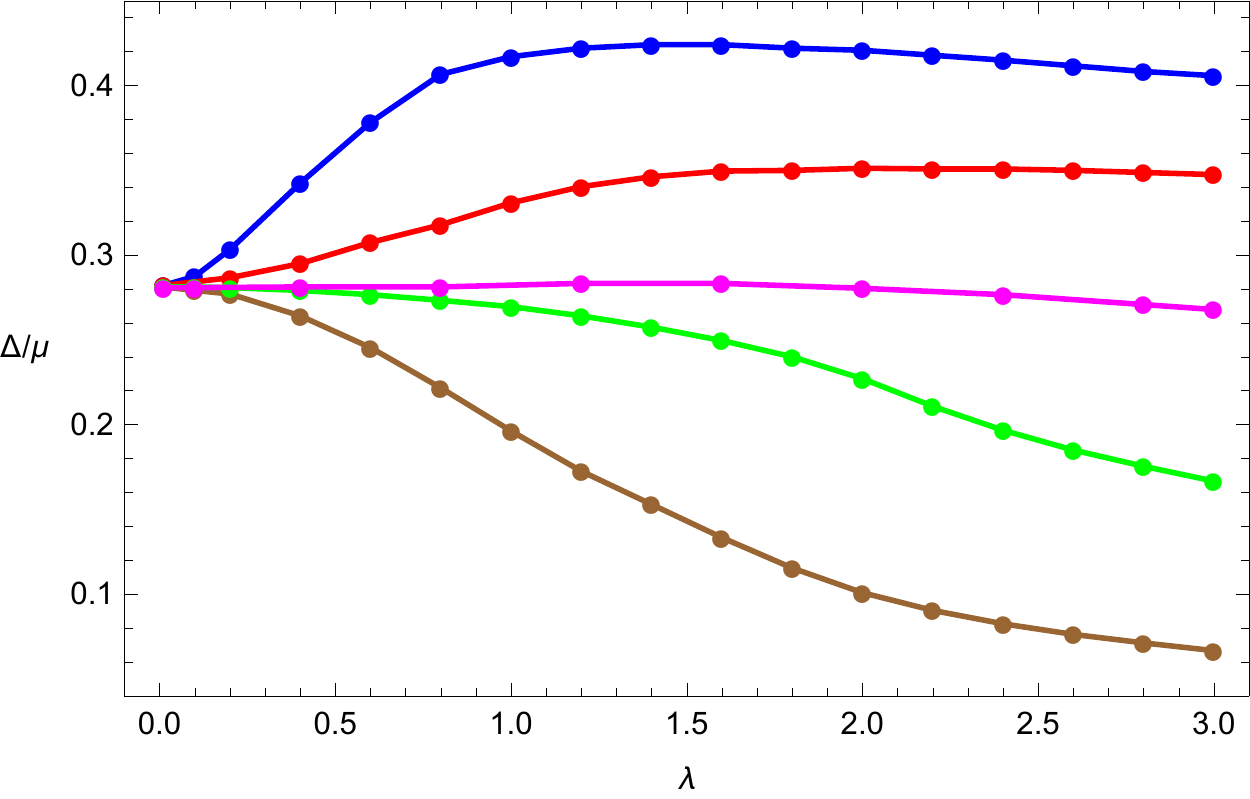}\hspace{0.5cm}
\caption{\label{widthp6DiLa}The width $\Delta/\mu$ of the gap as a function of $\lambda$.
Left plot is for $k=0.3$ and different $p$ (blue line for $p=6$, red for $p=6.5$ and brown for $p=7$)
and right plot is for $p=6$ and different $k$
(blue line for $k=0.03$, red for $k=0.3$, magenta for $k=0.38$, Green for $k=0.8$ and brown for $k=2$).
}}
\end{figure}

We also study the effects on the width of the gap of wave-number
$k$ and lattice amplitude $\lambda$, which can be summarized as follows.
\begin{itemize}
  \item First of all, a turning point is observed in the plot of gap width $\Delta/\mu$ versus $k$,which is shown in FIG.\ref{width_k_lambda}. Specifically, as $k$
increases from small one,
  the width of the gap $\Delta/\mu$ will decrease at first,
  which coincides with the idea of Mott that the
  strength of electron-electron correlations increases with the
  decrease of lattice constant \cite{Mott1,Mott2,Mott3,Mott4,Mott5}.
  However, when the wave-number exceeds a critical value $k_c$,
  $\Delta/\mu$ rises up slightly. Interestingly enough, we find
  the values of $k_c$ falls into the critical
regime of MIT in the phase diagram of the
background, which is quite general. We expect to understand
  this anomaly with an analytical treatment in near future.
  \item The critical value $k_c$ is almost independent of the dipole coupling $p$ but becomes larger with the increase of the
  lattice amplitude, as shown in FIG.\ref{width_k_lambda} and Table
  \ref{Tablekc}.
  In addition, it is interesting to notice that in the right plot of Fig.\ref{width_k_lambda} all the lines almost intersect at one point of
  $k\simeq0.38$.
  \item The width of gap as a function of the lattice amplitude $\lambda$ is plotted in
  Fig.\ref{widthp6DiLa}.
  For $k<0.38$, with the augmentation of the lattice amplitude $\lambda$,
  the width of the gap $\Delta/\mu$ rises up at first and then tends to a constant
  or very slightly decreases, depending on the dipole coupling $p$.
  But for $k>0.38$ the cases are just opposite.
  \item As a consistent check,
  one may find that in the right plot in FIG.\ref{widthp6DiLa},
  all the lines converge to a point at $\lambda\rightarrow 0$,
  which just corresponds to the gap width in RN-AdS black hole background.
\end{itemize}

\begin{widetext}
\begin{table}[ht]
\begin{center}
\begin{tabular}{|c|c|c|c|c|c|c|}
         \hline
~$\lambda$~ &~$0.1$~&~$0.5$~&~$0.8$~&~$1$~&~$2$~&~$3$~
          \\
        \hline
~$k_c$~ & ~$0.5$~ & ~$0.6$~ & ~$0.7$~& ~$0.8$~&~$0.9$~&~$1$~
          \\
        \hline
\end{tabular}
\caption{\label{Tablekc}The turning point $k_c$ for sample values of $\lambda$ and fixed $p=6$.}
\end{center}
\end{table}
\end{widetext}

\subsection{The dynamics at different temperatures}\label{ST}

So far, we have only focused on the system at a very low
temperature $T\simeq 0.00398$. Now, we turn to study the evolution
of the spectral function with temperature. The
temperature dynamics is one of important aspects of Mott
insulators. A characteristic quantity is the ratio of the
zero-temperature gap $\Delta$ to the critical temperature
$T_{\ast}$, at which the gap closes. A typical example is
transition-metal oxides $VO_2$, for which the ratio
$\Delta/T_{\ast}$ is approximately 20. Compared with the case of
superconductor, in which the $U(1)$ symmetry is spontaneously
broken and $\Delta/T_c\simeq 1-2$, the ratio $\Delta/T_{\ast}$ is
well beyond unity, which is one of the unresolved puzzles with
$VO_2$. It indicates that the driven force to form the gap in Mott
transition is strongly correlated. By holography, the authors in
\cite{Phillips:PRD} show the ratio $\Delta/T_{\ast}$ in RN-AdS
black hole is approximately $10$ for $p=6$ (or $p=7$), which
suggests holography can provide a good description on strong
correlation induced transition. Here we shall address this issue
on Q-lattice by holography.

\begin{figure}
\center{
\includegraphics[scale=0.22]{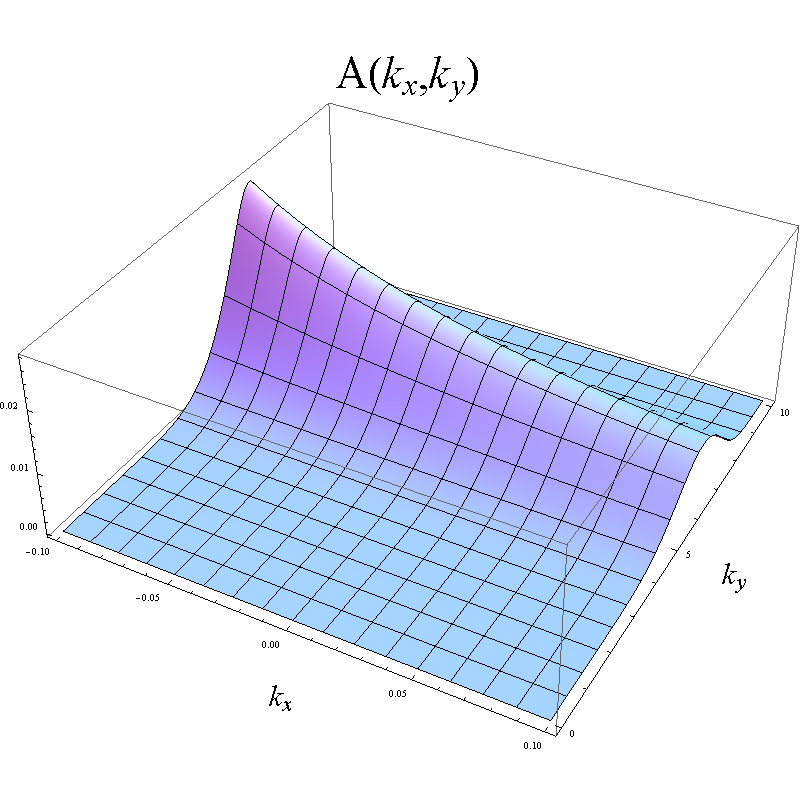}\hspace{0.5cm}
\includegraphics[scale=0.3]{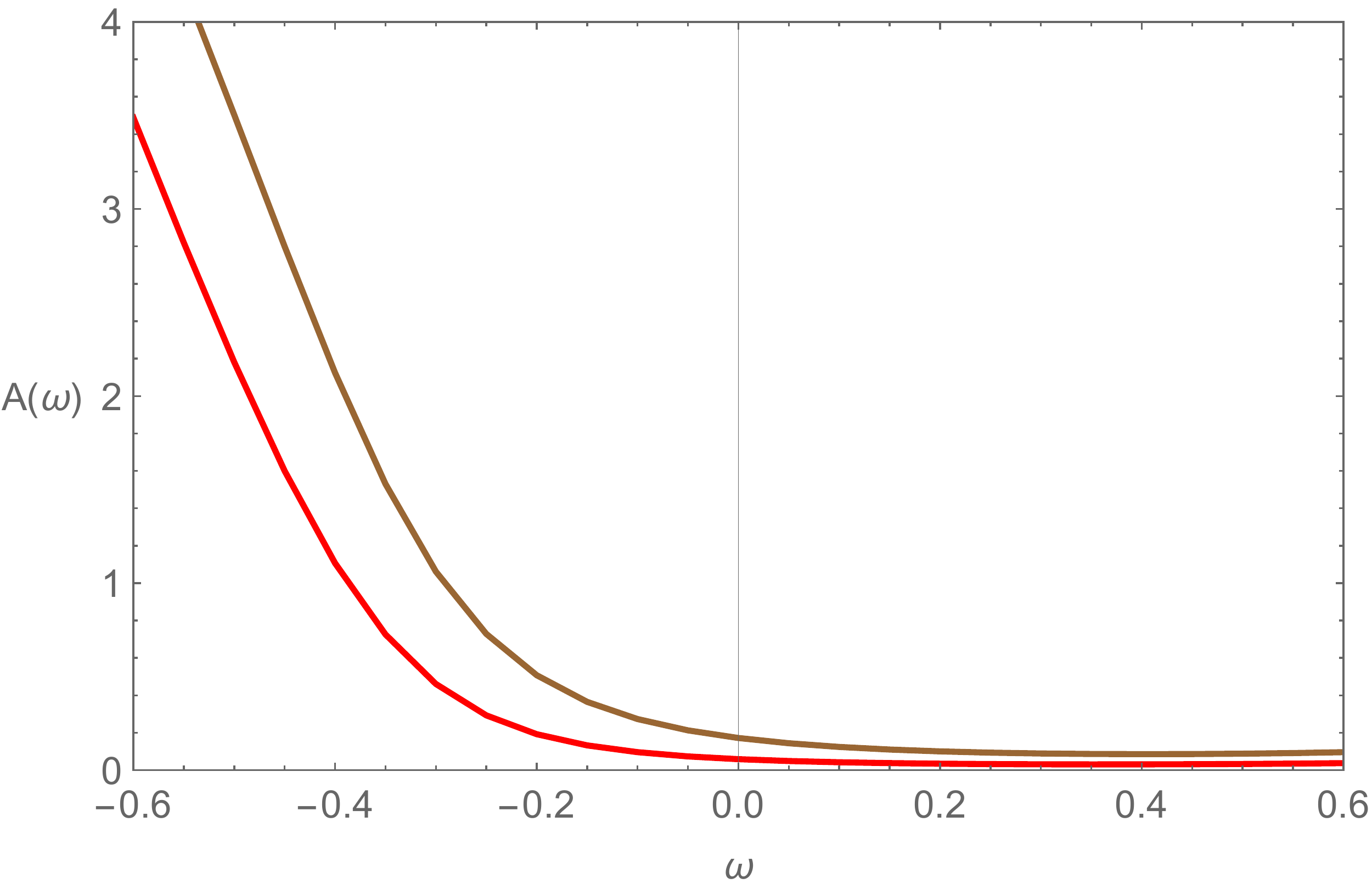}\hspace{0.5cm}
\caption{\label{D3DOST}Left plot: The 3D plot for $T\simeq0.115$. Right plot: The DOS for different $T$ (red for $T\simeq 0.115$ and brown for $T\simeq 0.151$).
Here, $p=6$, $\lambda=1$ and $k=0.3$.}}
\end{figure}

\begin{widetext}
\begin{table}[ht]
\begin{center}
\begin{tabular}{|c|c|c|c|c|c|c|c|}
         \hline
~$(\lambda,k)$~ &~$(1,0.1)$~&~$(1,0.75)$~&~$(1,2)$~&~$(0.1,0.3)$~&~$(1,0.3)$~&~$(3,0.3)$~
          \\
        \hline
~$\Delta/T_{\ast}$~ & ~$10.93$~ & ~$8.86$~ & ~$10.24$~&~$9.79$~&~$10.33$~&~$10.07$~
          \\
        \hline
\end{tabular}
\caption{\label{Tablev3}The ratio $\Delta/T_{\ast}$ for different lattice amplitude $\lambda$ and wave-number $k$.}
\end{center}
\end{table}
\end{widetext}

Before working out the results of the ratio $\Delta/T_{\ast}$, we
first make a qualitative description. FIG.\ref{D3DOST} shows a 3D
plot of the spectral function at $T\simeq0.115$ and the DOS for different $T$ with $\lambda=1$ and $k=0.3$. It is
obvious that the gap closes when the temperature exceeds some
certain value. Now, we work out the ratio $\Delta/T_{\ast}$ for
different lattice amplitudes $\lambda$ and wave-numbers $k$ for
$p=6$ (Table \ref{Tablev3}) \footnote{Because we can not cool our
system down to zero temperature, here we take the width of gap
$\Delta$ at $T\simeq 0.00398$ instead of zero temperature gap.}.
From Table \ref{Tablev3} we can see that most values of the ratio
$\Delta/T_{\ast}$ are approximately $10$. Especially, in the deep
insulating background, the value of $\Delta/T_{\ast}$ is larger
than others, approaching $11$. In summary, in our holographic
fermionic system with dipole coupling on Q-lattice, the
temperature dynamics possesses non-trivial behavior, as revealed in
RN-AdS black hole \cite{Phillips:PRD}.

\subsection{Fermi surface and gap on anisotropic Q-lattice}\label{anisotropic}

\begin{figure}
\center{
\includegraphics[scale=0.25]{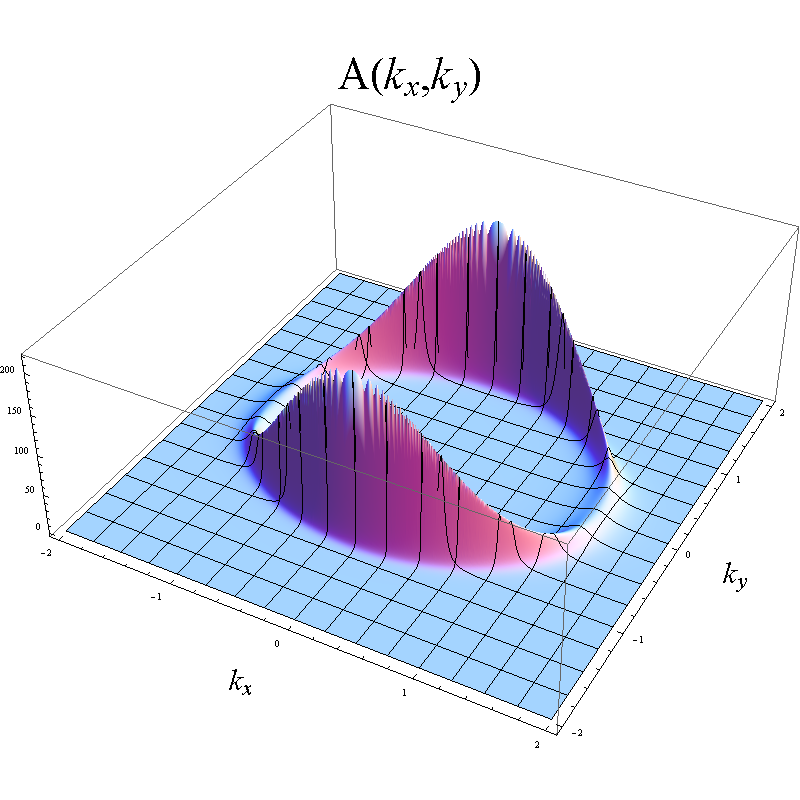}\hspace{0.5cm}
\includegraphics[scale=0.25]{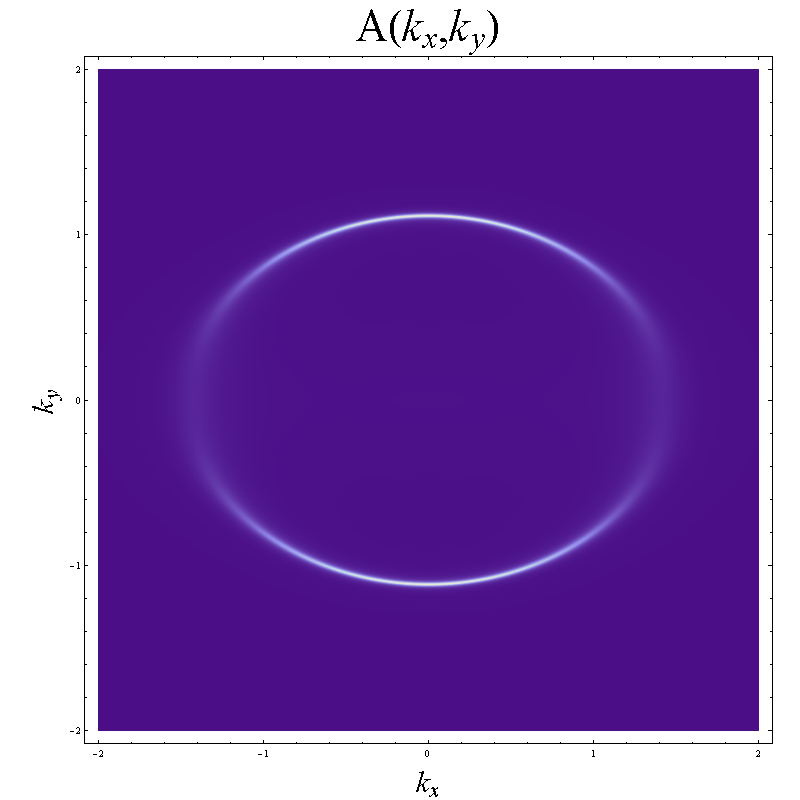}\hspace{0.2cm}
\caption{\label{Anisotropic_p0}3D and density plot of spectral function $A(k_x,k_y,\omega=0)$ on anisotropic Q-lattice for $p=0$. Here, $q=1$, $\lambda_1=2$, $\lambda_2=0.1$ and $k_1=k_2=0.8$.}}
\end{figure}
\begin{figure}
\center{
\includegraphics[scale=0.25]{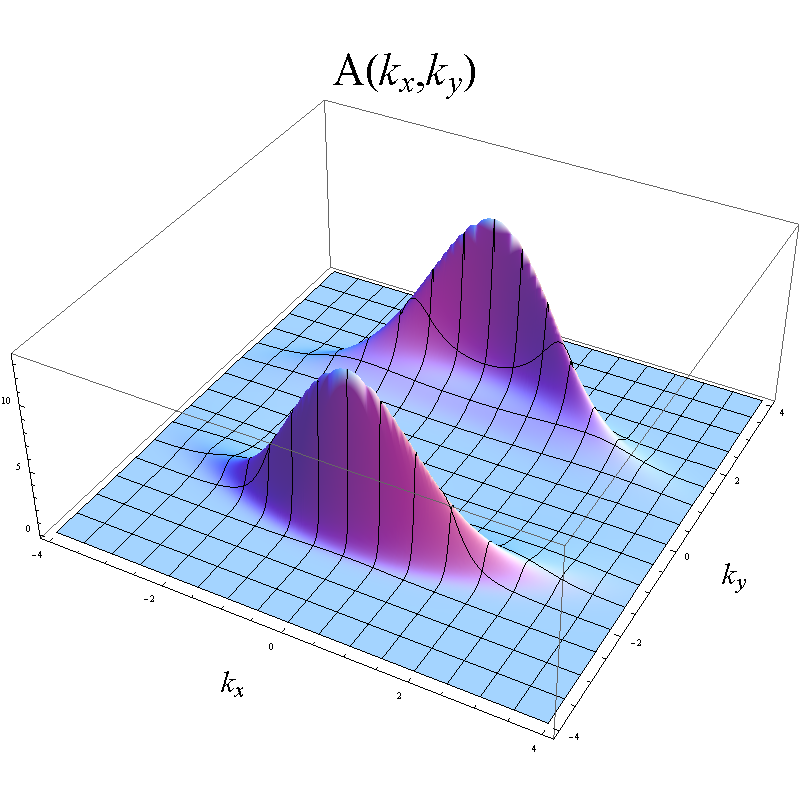}\hspace{0.5cm}
\includegraphics[scale=0.25]{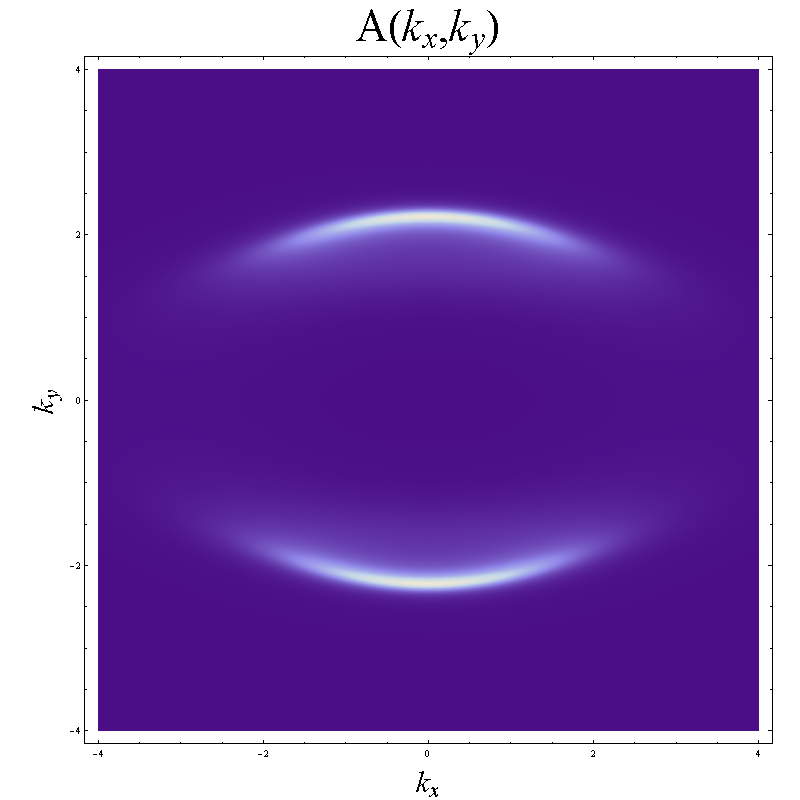}\hspace{0.2cm}
\caption{\label{Anisotropic_p1}3D and density plot of spectral function $A(k_x,k_y,\omega=0)$ on anisotropic Q-lattice for $p=1$. Here, $q=1$, $\lambda_1=2$, $\lambda_2=0.1$ and $k_1=k_2=0.8$.}}
\end{figure}

In this subsection, we briefly discuss the Fermi surface and gap
on anisotropic Q-lattice. As shown in
FIG.\ref{Anisotropic_p0}, if we set $q=1$, $\lambda_1=2$,
$\lambda_2=0.1$ and $k_1=k_2=0.8$ for $p=0$, we find that the
Fermi peaks along $k_x$ direction develop into some bumps. It
exhibits an anisotropic peaks with different magnitudes, which suggests
that insulating phase arises in one direction while metallic in
the other. Furthermore, when we switch on the dipole coupling $p$,
the gap along $k_x$ direction gradually opens but the bump still
remain along $k_y$ direction (FIG.\ref{Anisotropic_p1}).
Finally, we would like to comment that the same statement is true for the conductivity on Q-lattice.
For instance, since the lattice is introduced in just one direction, the Q-lattice geometries
are highly anisotropic in \cite{Donos:QLattice}, in which the calculation of the conductivity
reveals that they can be insulators in the direction where the lattice is placed whilst remaining a pure metal in
the other direction where the lattice is absence and the translational invariance is reserved.
Therefore, the anisotropic geometry
with insulating phase in one direction but metallic in the other
provides us more space to model real materials with
anisotropy by holography.

\section{Conclusion and discussion}\label{SConclusion}

Our main results in this paper are:
\begin{itemize}
  \item Two key features of doped Mott physics, the dynamical emergence of
a gap and the spectral weight transfer, are observed in the
 Q-lattice background with dipole coupling.
  Such features have been observed in many of holographic fermionic systems including dipole coupling, e.g., RN-AdS geometry and other geometry.
  It confirms the robustness of the generations of Mott gap induced by dipole coupling, which play the double roles of doping as well as the interaction strength $U$ in the Hubbard model.
  \item The most important thing is that the fermionic system with dipole coupling on Q-lattice can exhibit abundant Mott physics due to the introduction of wave-number $k$
  and lattice amplitude $\lambda$.
  \item The evolution of the spectral function as a function of temperature reveals the non-trivial temperature dynamical behavior,
  which indicates the Mott transition induced by dipole coupling on Q-lattice is
 due to strong correlations but not the spontaneous breaking
 of some symmetry.
  \item For the free fermionic system, the wave-number $k$ and the lattice amplitude $\lambda$ can not generate the Mott gap.
  When dipole coupling $p$ exceeds some critical value, the gap opens and, usually, the gap opens much easily in deep insulating phase.
  \item The anisotropic peaks with different magnitudes occurs in our holographic fermionic system on Q-lattice.
  It indicates that insulating and metallic phases arise in different directions, respectively.
\end{itemize}

Still, a number of problems deserve further
exploration.
\begin{itemize}
  \item Rather than the dipole coupling, we may introduce other sorts of couplings to induce the Mott transition. Especially
  the interaction of electron-phonon as $\bar{\zeta}(\eta_1\phi_1+\eta_2\phi_2)\zeta+h.c$ may have a manifest periodic structure and generate Brillouin zones. Our work on this subject is under progress.
  \item By the poles and zeros duality through the $det G_R$, the pseudo-gap phase can be observed in the holographic fermions with dipole coupling in RN-AdS geometry\cite{AlsupDuality}.
  It is valuable to address this problem in the holographic Q-lattice model.
  \item It is interesting to study the Fermi arc phenomenology in the anisotropic Q-lattice model.
\end{itemize}

\begin{acknowledgments}

We are grateful to the anonymous referee for valuable comments.
This work is supported by the Natural
Science Foundation of China under Grant Nos.11275208, 11305018 and
11178002. Y.L. also acknowledges the support from Jiangxi young
scientists (JingGang Star) program and 555 talent project of
Jiangxi Province. J. P. Wu is also supported by Program for Liaoning Excellent Talents in University (No. LJQ2014123).

\end{acknowledgments}

\end{document}